\documentclass[aps,pra,twocolumn,showpacs,notitlepage,nofootinbib,floatfix]{revtex4-1}
\usepackage{graphicx}
\usepackage{amsmath,amsfonts}
\usepackage{amssymb}
\usepackage{color}
\usepackage[T1]{fontenc}
\usepackage{hyperref}
\usepackage[colorinlistoftodos]{todonotes}
\usepackage{tabularx} 

\begin{document}
\title{Dynamic hysteresis from bistability in an antiferromagnetic spinor condensate}
\author{Tomasz \'Swis\l{}ocki$^{1}$, Andrzej Zembrzuski$^{1}$, Micha\l{} Matuszewski$\,^{2}$,  Emilia Witkowska$\,^{2}$}
\affiliation{$^{1}$Faculty of Applied Informatics and Mathematics, Warsaw University of Life Sciences, ul. Nowoursynowska 159, 02-786 Warsaw, Poland\\
$^{2}$Institute of Physics, Polish Academy of Sciences, Aleja Lotnikow 32/46, PL-02668 Warsaw, Poland}
\begin{abstract}
We study the emergence of hysteresis during the process of quantum phase transition from an antiferromagnetic to a phase-separated state in a spin-1 Bose Einstein condensate of ultracold atoms. We explicitly demonstrate the appearance of a hysteresis loop with various quench times showing that it is rate-independent for large magnetizations only. In other cases scaling of the hysteresis loop area is observed, which we explain by using the Kibble-Zurek theory in the limit of small magnetization. The effect of an external harmonic trapping potential is also discussed.
\end{abstract}

\maketitle

The classic example of hysteresis is the relation of the applied magnetic field to the magnetization in solid-state ferromagnetic materials. Hysteresis can also occur in different situations as a product of a fundamental physical mechanism like a phase transition, or a result of imperfections or degradations. Hysteresis occurs in two forms: rate-dependent and rate-independent. In the rate-independent case, two or more metastable energy states are separated by an energy barrier. When an external driving force moves the system from one metastable state to another, the system exhibits the history-dependent behavior. The rate-independent hysteresis of supercurrent in a rotating, superfluid Bose-Einstein condensate was observed and has been proclaimed as a milestone in the advancement of atomtronic circuitry~\cite{{Campbell2014},{Kavoulakis2015},{Davis2014}}. A recent experiment~\cite{Fattori2016} has also demonstrated the rate-independent hysteresis when a Bose-Einstein condensate is placed in a double-well potential. On the other hand, the observation of rate-dependent hysteresis could provide insight into the out-of-equilibrium dynamics of the system.

Spinor condensates are composed of $N$ atoms in several Zeeman components with a given hyperfine spin $F$ and magnetic numbers $m_F$. 
The global ground state of the $F=1$ system is classified as ferro- or antiferromagnetic, depending on the sign of spin-dependent interactions. The magnetization longitudinal with respect to magnetic field $M$ is approximately conserved in the system and acts as an independent external parameter. This conservation law comes from the spin rotational symmetry of contact interactions when dipole-dipole interactions are neglected. Consequently, in contrast to solid-state magnetic materials, classical hysteresis is impossible in spinor $F=1$ condensates. 
However, a weak magnetic field drives the system to the transition from an antiferromagnetic ground state to a state where domains of atoms with different spin projections separate~\cite{Matuszewski_PS}. The phase transition is specific due to the region of bistability in which the antiferromagnetic and phase separated states are both metastable~\cite{ourPRB}.  

In this paper, we investigate the emergence of hysteresis from bistability during the phase transition in an antiferromagnetic spin-1 condensate. 
The system is already recognized as useful for quantum technology tasks, however hysteresis was not been examined up to now.
By employing numerical simulations within the truncated Wigner approximation, we demonstrate the appearance of rate-independent hysteresis for large magnetizations only, when the bistability region is widest. 
In the limit of small magnetizations the bistability region disapears, however competition between the characteristic time scale of driving and the relaxation time of the system leads to emergence of rate-dependent hysteresis. 
We show that in this case, the hysteresis loop area is subject to a universal scaling law. We estimate the scaling of the hysteresis loop area based on the Kibble-Zurek (KZ) theory~\cite{Zurek1,Zurek2,Zurek3}, similarly as in~\cite{ciuti2016}. 
The situation changes when the system is enclosed in an external trapping potential, as it makes the separation of the two metastable energy states smeared out, and the rate-independent hysteresis loop becomes impossible to observe. 
The scaling of the rate-dependent hysteresis area in this case is influenced by the trap inhomogeneity \cite{ourPRAtrap, Zurek2009, Dziarmaga2010, Sabbatini2012,delCampo2013,Saito2013,delCampo2014}. In addition, in the low density regime a process of phase ordering kinetics~\cite{Bray1994, Blakie2017,Moore2007, Kawaguchi2013, Kawaguchi2015,Blakie2016} additionally modifies the scaling laws.
Finally, we propose an experimental setup and parameters reachable by current technologies
\cite{Gerbier2012, Gerbier2017, Shin2017}
which will enable to observe the clear rate-independent hysteresis loop of non-zero width.

The system we focus on is an antiferromagnetic condensate of sodium atoms in a homogeneous magnetic field $B$, having positive magnetization such that $0<M<N$. 
We restricted the model to one dimension, with the other degrees of freedom confined by a strong transverse potential with frequency $\omega_\perp$.
The model Hamiltonian of the system is composed of two terms: the energy of the spin-1 system $H_s$ and the energy shift $H_{\rm QZE}$ due to a homogeneous magnetic field. The first term is given by
\begin{eqnarray} \label{En}
H_s &=& \int d x \sum_{m_F} \psi_{m_F}^\dagger \left(-\frac{\hbar^2}{2\mu}\nabla^{2} + \frac{1}{2}\mu \omega^2x^2  \right) \psi_{m_F} \nonumber \\ 
{}&+&  \int d x \left(  \frac{c_0}{2} n^2 +  \frac{c_2}{2}  {\bf F}^2 \right) ,
\end{eqnarray}
where $\mu$ is the atomic mass, 
$\omega$ is a trap frequency,
$n=\sum n_{m_F}=\sum\psi_{m_F}^\dagger \psi_{m_F}$ is the local atom density, 
and ${\bf F}=(\psi^{\dagger}f_x\psi,\psi^{\dagger}f_y\psi,\psi^{\dagger}f_z\psi)$ is the spin density, where $f_{x,y,z}$ are the spin-1 matrices and $\psi^T =(\psi_1,\psi_0,\psi_{-1})$.
The spin-independent and spin-dependent interaction coefficients are 
$c_0=2 \hbar \omega_\perp (2 a_2 + a_0)/3$ and $c_2= 2 \hbar \omega_\perp (a_2 - a_0)/3$, 
where $a_S$ is the s-wave scattering length for colliding atoms with total spin $S$. The coefficients $c_{0}$ and $c_{2}$ are both positive for sodium atoms.
The linear Zeeman effect becomes irrelevant, since it is proportional to the conserved magnetization, while the quadratic Zeeman energy becomes essential
\begin{equation}
H_{\rm QZE}=- q\, c_2 \rho \, N_0,
\end{equation}
where we dropped a constant term. Here, 
$N_0$ is the number of atoms in the $m_F=0$ Zeeman component,
$\rho=N/L$ is the total density, $L$ is the system size, 
and $q=AB^2/(c_2\rho)$ with $A=(g_I + g_J)^2 \mu_B^2/16 E_{\rm HFS}$ in which
$g_I$ and $g_J$ are the gyromagnetic ratios of the electron and nucleus, $\mu_B$ is the Bohr magneton, $E_{\rm HFS}$ is the hyperfine energy splitting. The value and sign of the quadratic Zeeman energy, through $q $, can be controlled using the magnetic field $B$ or the microwave dressing~\cite{Gerbier2006}.

\begin{figure}[]
\includegraphics[width=0.5\textwidth]{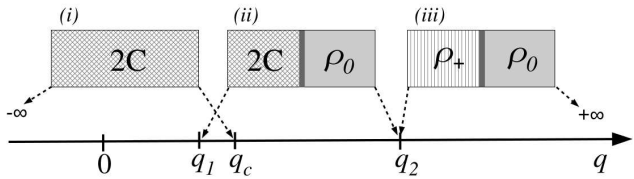}
\caption{Schematic structure of the mean-field ground state of the system versus $q$ for positive magnetization $M>0$. Dashed arrows indicate the stability regions of $(i)$ antiferromagnetic $2C$ state with atoms in the $m_F=\pm1$ components, $(ii)$ $2C+\rho_0$ state composed of two domains of the $2C$ and $\rho_0$  phases, the latter one is composed of atoms in the $m_F=0$ component, $(iii)$ $\rho_++\rho_0$ state composed of two domains of $\rho_0$ and $\rho_+$, where the latter one is composed of atoms in the $m_F=1$ component~\cite{Matuszewski_PS}. The vertical thick lines in $(ii)$ and $(iii)$ illustrate domain walls.}
\label{fig:fig1}
\end{figure}

\begin{figure}[]
\includegraphics[width=0.45\textwidth]{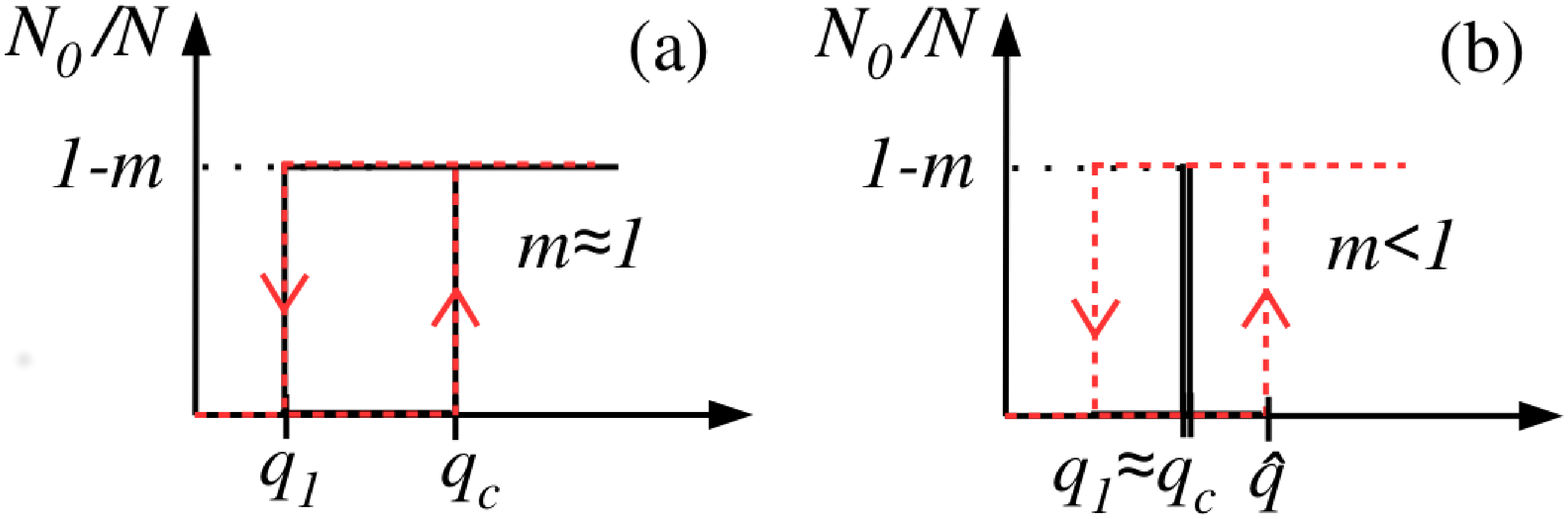}
\caption{Hysteresis in the system while crossing the critical points in the case of (a) large and (b) small width of the bistability region. The hysteresis loop in the small quench times limit is illustrated by dashed red lines, and in the adiabatic limit by solid black lines.}
\label{fig:fig2}
\end{figure}

We first concentrate on the case of a homogeneous system ($\omega=0$).
The ground state structure of the uniform system can be found on the mean-field level by minimization of the mean-field energy functional in the subspace of fixed magnetization \cite{{Zhang2003},{Matuszewski_PS},{ourPRB}}.
When the spin healing length $\xi_s=\hbar/\sqrt{2 \mu c_2 \rho}$ is much smaller than the system size $L$, the structure of the system ground state is composed of three states divided by two critical points at $q_1=m^2/2$, where $m=M/N$ is the fractional magnetization, and $q_2=1/2$~\cite{foot1}, as illustrated in Fig.~\ref{fig:fig1}. 
The system is in the antiferromagnetic ground state when $q< q_1$, and in the phase separated state otherwise. Moreover, the analysis of the Bogoliubov spectrum \cite{ourPRB} shows that the antiferromagnetic state is dynamically stable, and it remains a local energy minimum up to the value $q_c=1-\sqrt{1-m^2}$.
It is easy to see that $q_1\le q_c$ for any $m$.
A simultaneous stability of the two states may lead to the hysteresis phenomenon when dynamically changing $q$. We assume that the parameter $q$ is tuned in the following way:
\begin{equation}\label{eq:q}
q(t) = \left\{
\begin{array}{ccc}
\alpha \frac{t}{\tau_Q}, & \,\,\,\,& t \in(0, \tau_Q), \\
\alpha \left( 2-\frac{t}{\tau_Q} \right), & \,\,\,\,& t \in (\tau_Q, t_{\rm max}), 
\end{array}
\right.
\end{equation}
where $\alpha$ sets the maximal value of $q$, $\tau_Q$ is the quench time and $t_{\rm max}$ is the evolution time.
The nucleation and growth of the $\rho_0$ phase can be characterised by the fraction of atoms in the $m_F=0$ component. The relation of $N_0/N$ versus $q$ can take the form of the hysteresis loop of width $q_c-q_1$ set by the size of the bistability region. The width of the bistability region depends on the fractional magnetization $m$ making the hysteresis phenomena qualitatively diffrent in the two limits: $m\to 0$ and $m\to 1$, as illustrated in Fig.~\ref{fig:fig2}. 

In the limit of macroscopic magnetization $m\to 1$, the region of bistability is large $q_c-q_1 \to 1/2$ and the hysteresis may become rate-independent. The hysteresis area
\begin{equation}\label{eq:O}
S(\tau_Q)=\int_{q(\tau_Q)}^{q(t_{\rm max})} \frac{N_0(q)}{N}\, {\rm d}q - \int_{q(0)}^{q(\tau_Q)} \frac{N_0(q)}{N} \, {\rm d}q 
\end{equation}
is
\begin{equation}
S (\tau_Q) \approx (1-m)(q_c-q_1)
\end{equation} 
while approximating the shape of the hysteresis loop by a rectangle of height $N_0/N\to 1-m$ and width $q_c-q_1$, as represented in Fig.~\ref{fig:fig2}(a).

In the limit of small magnetization $m\to 0$ one can expect rate-dependent hysteresis as $q_c-q_1\to 0$. The scaling of the histeresis area (\ref{eq:O}) with the quench time $\tau_Q$ may exhibit a scaling law due to a non-adiabatic phase transition caused by finite quench times considered (\ref{eq:q}). In order to predict the corresponding scaling law we use the KZ theory~\cite{Zurek1,Zurek2,Zurek3} for the description of the non-adiabatic phase transition, which we have developed for the case of antiferromagnetic spinor condensates~\cite{ourPRL,ourPRA}. 
The KZ theory is a powerful tool which allows one to predict the scaling law for density of topological defects versus the quench rate based on the relation of characteristic time scales in terms of critical exponents, which are $z=1$ and $\nu=1/2$ for our system~\cite{ourPRB}. The theory is based on the adiabatic-impulse-adiabatic approximation, which implies that the scaling law is determined at the freeze-out time $\hat{t}$ when the dynamics of the system ceases to be adiabatic. 
The small parameter of the KZ theory is the distance from the critical point $\varepsilon$ which in the case of our system is $\varepsilon=q-q_c$. The KZ theory predicts $\varepsilon(\hat{t}) \propto \tau_Q^{1/(1+z\nu)}$ for the linear ramp we are considering.
The scaling law of the hysteresis area (\ref{eq:O}) is determined by the scaling of $q$ at $\hat{t}$, or consistently, by the distance from the critical point $\varepsilon(\hat{t})=q(\hat{t})-q_c$.
 By using our previous results for critical exponents \cite{ourPRB} one can show that the scaling law for the hysteresis area (\ref{eq:O}) at $\hat{t}$ is
\begin{equation}
\hat{S}(\tau_Q) \propto q(\hat{t})-q_c \propto \tau_Q^{-2/3}.
\end{equation}

We test the above prediction in numerical simulations within the truncated Wigner approximation \cite{WignerRef1,WignerRef2}. The dynamics of the system is then described by the set of time-dependent Gross-Pitaevskii (GP) equations
\begin{eqnarray} \label{Wigner}
i\hbar\frac{\partial \psi_0}{\partial t}  &=& \left(-\frac{\hbar^2 \nabla^2}{2\mu} + \frac{1}{2}\mu \omega^2x^2 + c_0 n - q(t)\, c_2 \rho \right)\psi_0 + \nonumber\\
      &+& c_2\left[(n_1+n_{-1})\psi_0+2\psi_0^*\psi_1\psi_{-1}\right],\nonumber\\
i\hbar\frac{\partial \psi_{\pm 1}}{\partial t}   &=& \left(-\frac{\hbar^2 \nabla^2}{2\mu} + \frac{1}{2}\mu \omega^2x^2 + c_0 n \right)\psi_{\pm 1} + \nonumber\\
      &+& c_2\left[(n_{\pm 1}- n_{\mp 1} + n_0)\psi_{\pm 1}+\psi_{\mp 1}^*\psi_0^2\right].
\end{eqnarray}
The initial state is the 2C state for $q=0$ such that in momentum space
$\psi_{m_F}(k, t=0)=\phi_{m_F} + \delta \phi_{m_F}$ with $|\phi_{m_F=\pm 1}|^2=(N \pm M)/2$, $|\phi_{m_F=0}|^2=0$ and stochastic noise $\langle\delta \phi^*_{m_F} (k) \delta \phi_{m_F'} (k') \rangle=\frac{1}{2}\delta_{m_F, m_F'} \delta_{k, k'}$ of $1/2$ particle per momentum mode in all three $m_F$ components is added. 

\begin{figure}[]
\begin{picture}(0,160)
\put(-130,80){\includegraphics[width=0.16\textwidth,height=0.19\textheight,angle=-90]{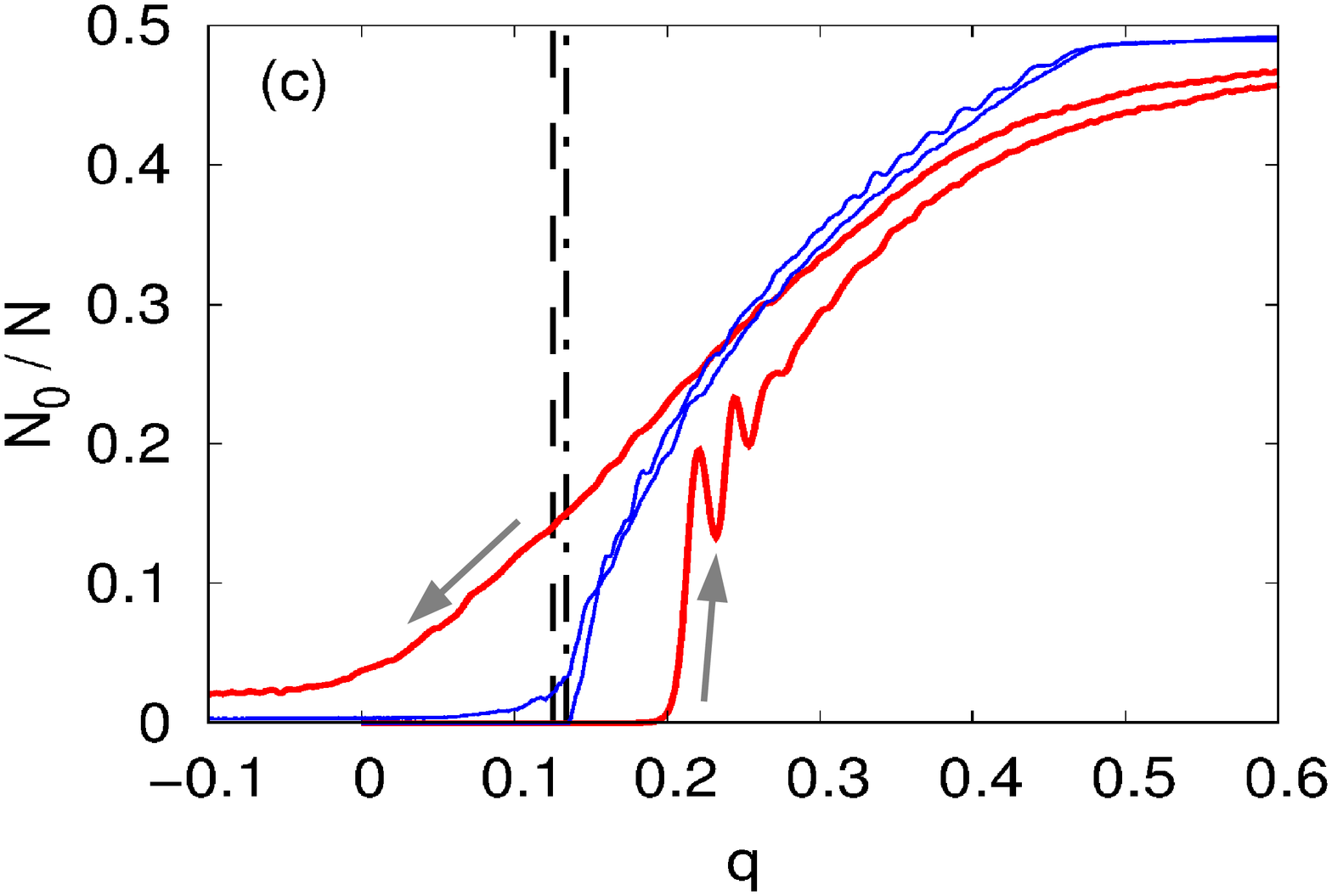}}
\put(-0,80){\includegraphics[width=0.16\textwidth,height=0.19\textheight,angle=-90]{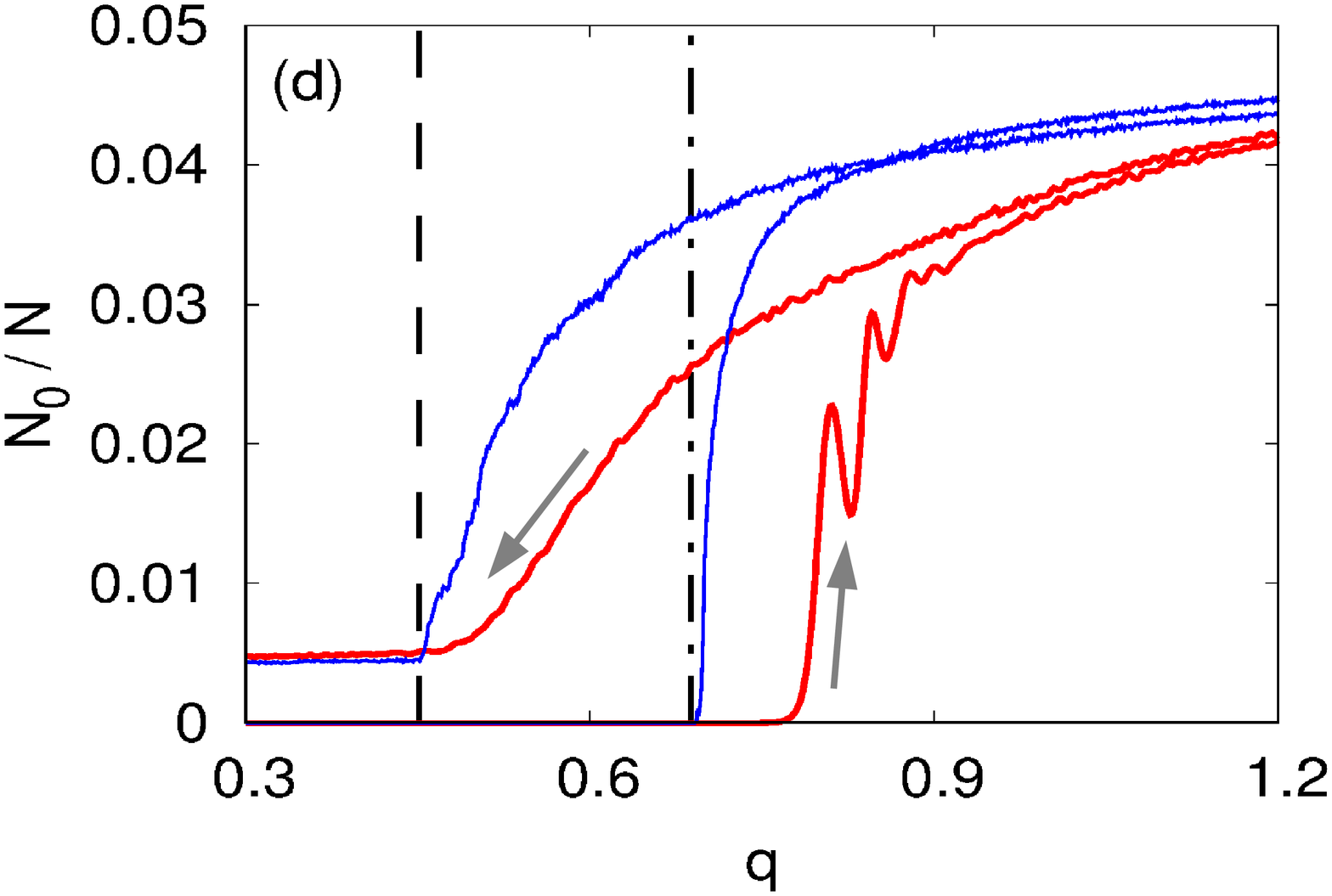}}
\put(-130,160){\includegraphics[width=0.16\textwidth,height=0.19\textheight,angle=-90]{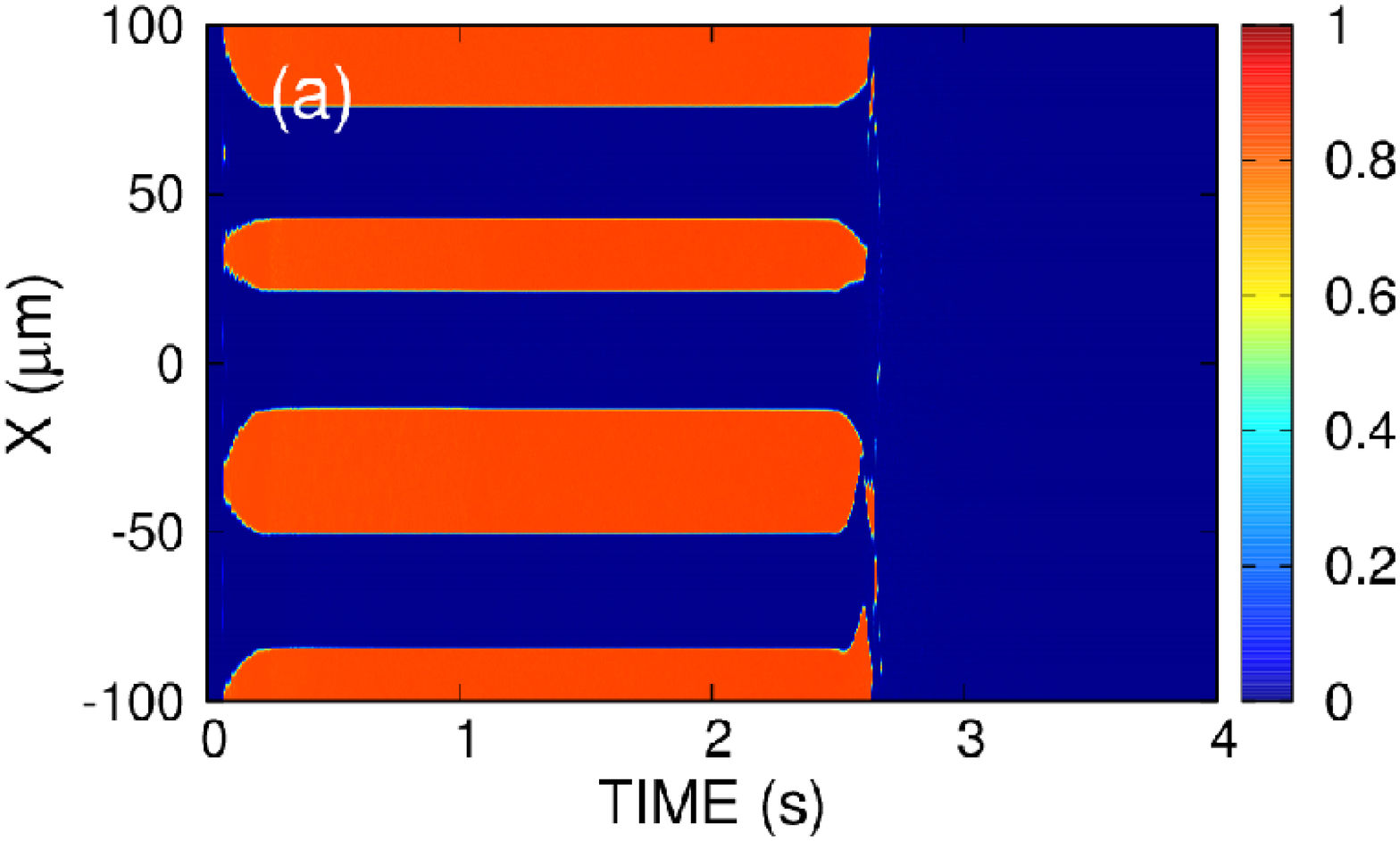}}
\put(-0,160){\includegraphics[width=0.16\textwidth,height=0.19\textheight,angle=-90]{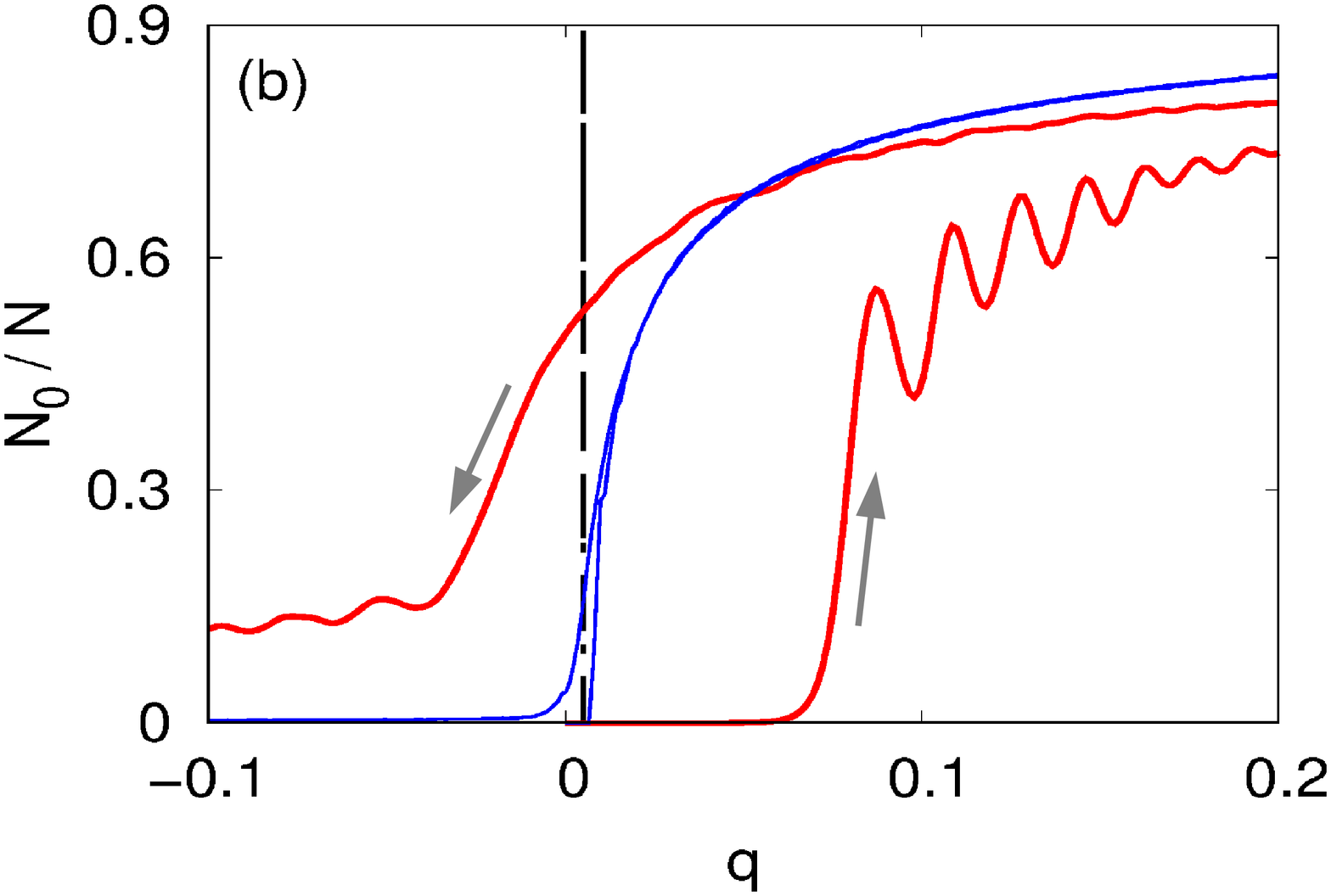}}
\end{picture}
\caption{
(Color online) (a) Evolution of density of the $m_F=0$ component for $\tau_Q=1.36$s and $m=0.5$. 
(b), (c) and (d) show evolution of $N_0/N$ as a function of $q$ averaged over $50$ realizations for a box-like potential ($\omega=0$) and $m=0.1$, $m=0.5$ and $m=0.95$ respectively. 
The thin red solid line corresponds to $\tau_Q=11$ms, while the blue thick one to $\tau_Q=1.36$s. $q_1$ (left) and $q_c$ (right) are denoted by dashed and dot-dashed black lines respectively. The gray arrows indicate direction of evolution. The parameters are as follows: $N=2\times 10^7$, $\omega_\perp=2\pi \times 1000\,$Hz, $L=200\, \mu$m, $\alpha=3$ and $t_{\rm max}=3\tau_Q$. 
}
\label{fig:fig3}
\end{figure}

\begin{figure}[]
\begin{picture}(0,160)
\put(-130,80){\includegraphics[width=0.16\textwidth,height=0.19\textheight,angle=-90]{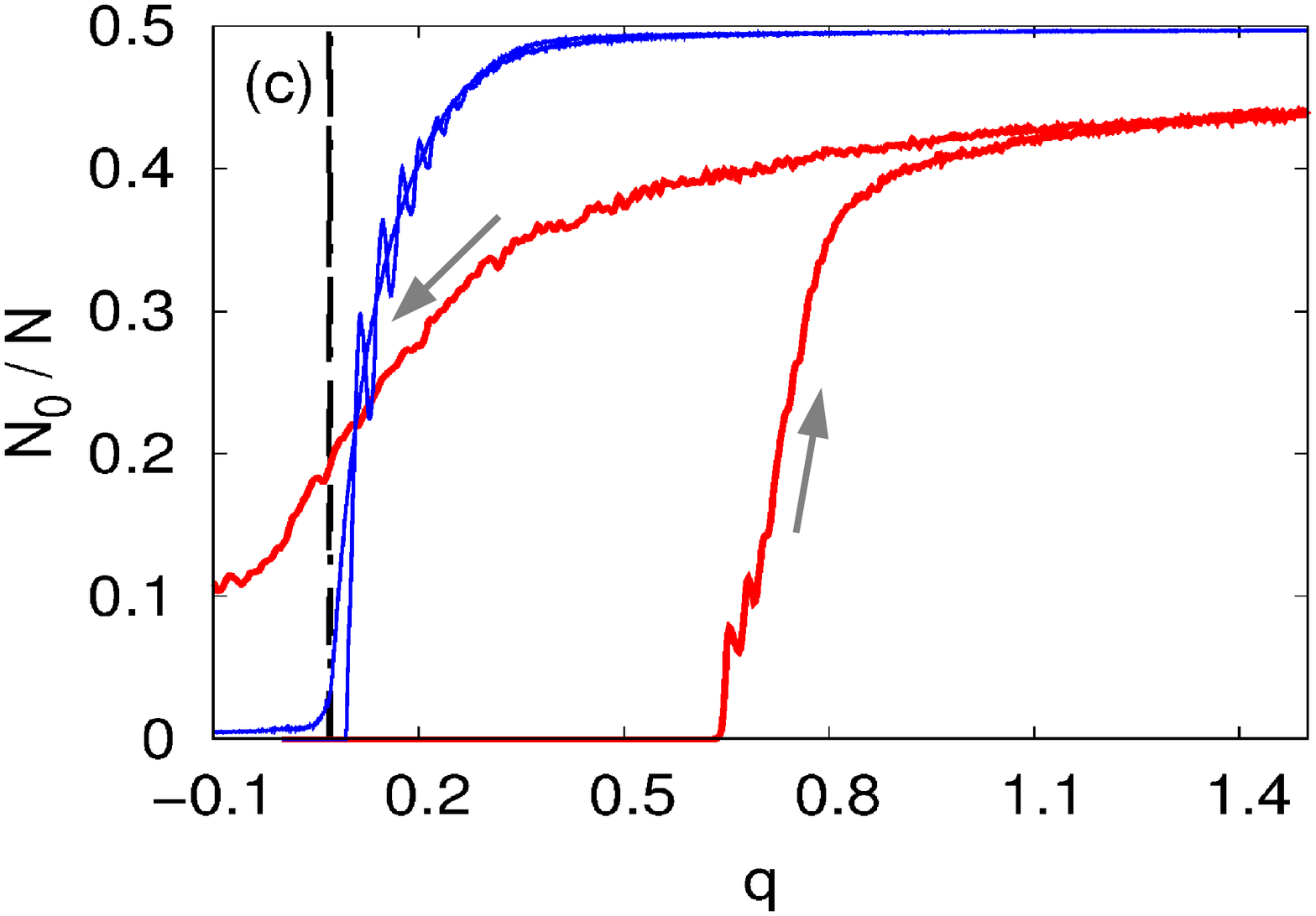}}
\put(-0,80){\includegraphics[width=0.16\textwidth,height=0.19\textheight,angle=-90]{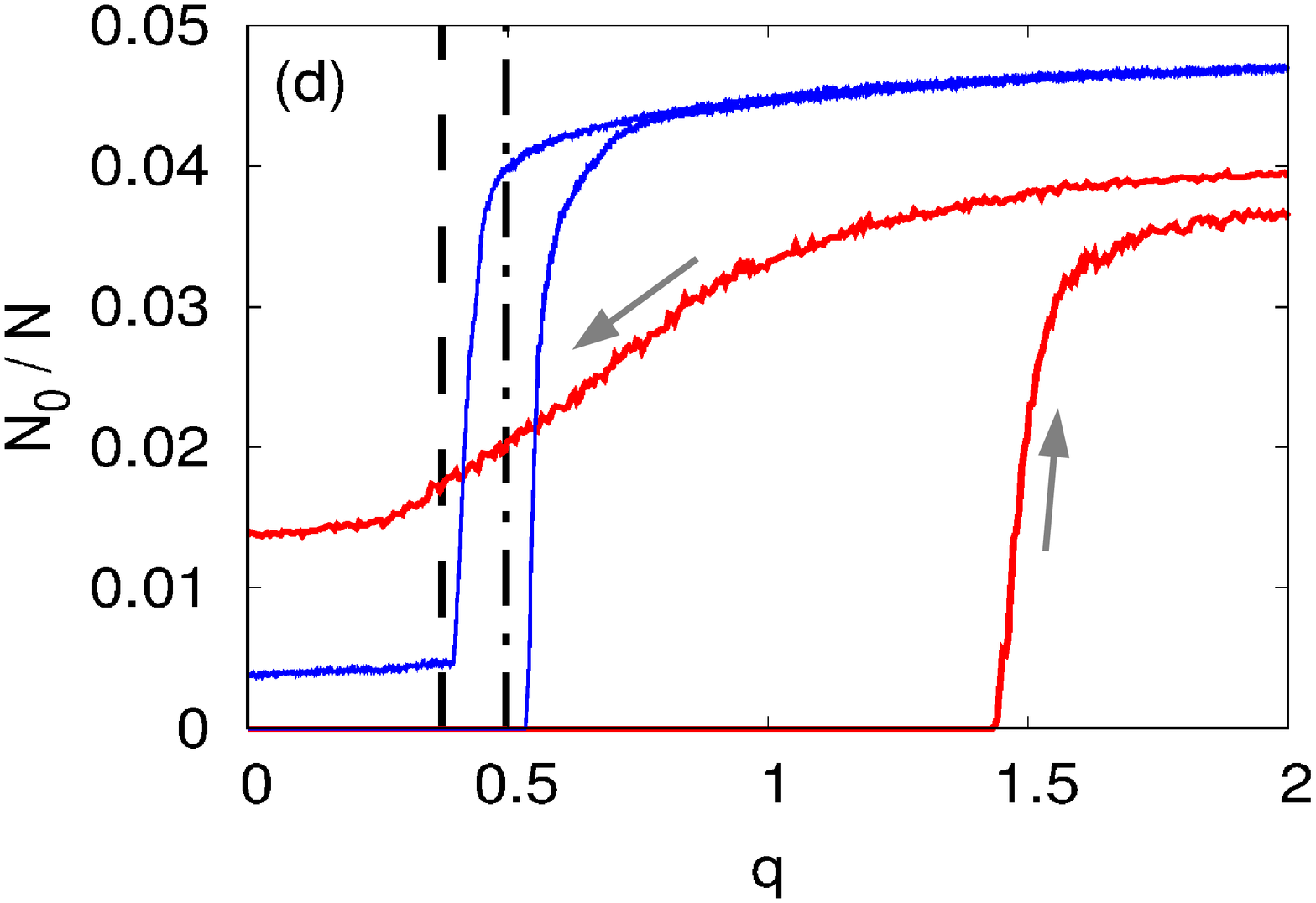}}
\put(-130,160){\includegraphics[width=0.16\textwidth,height=0.19\textheight,angle=-90]{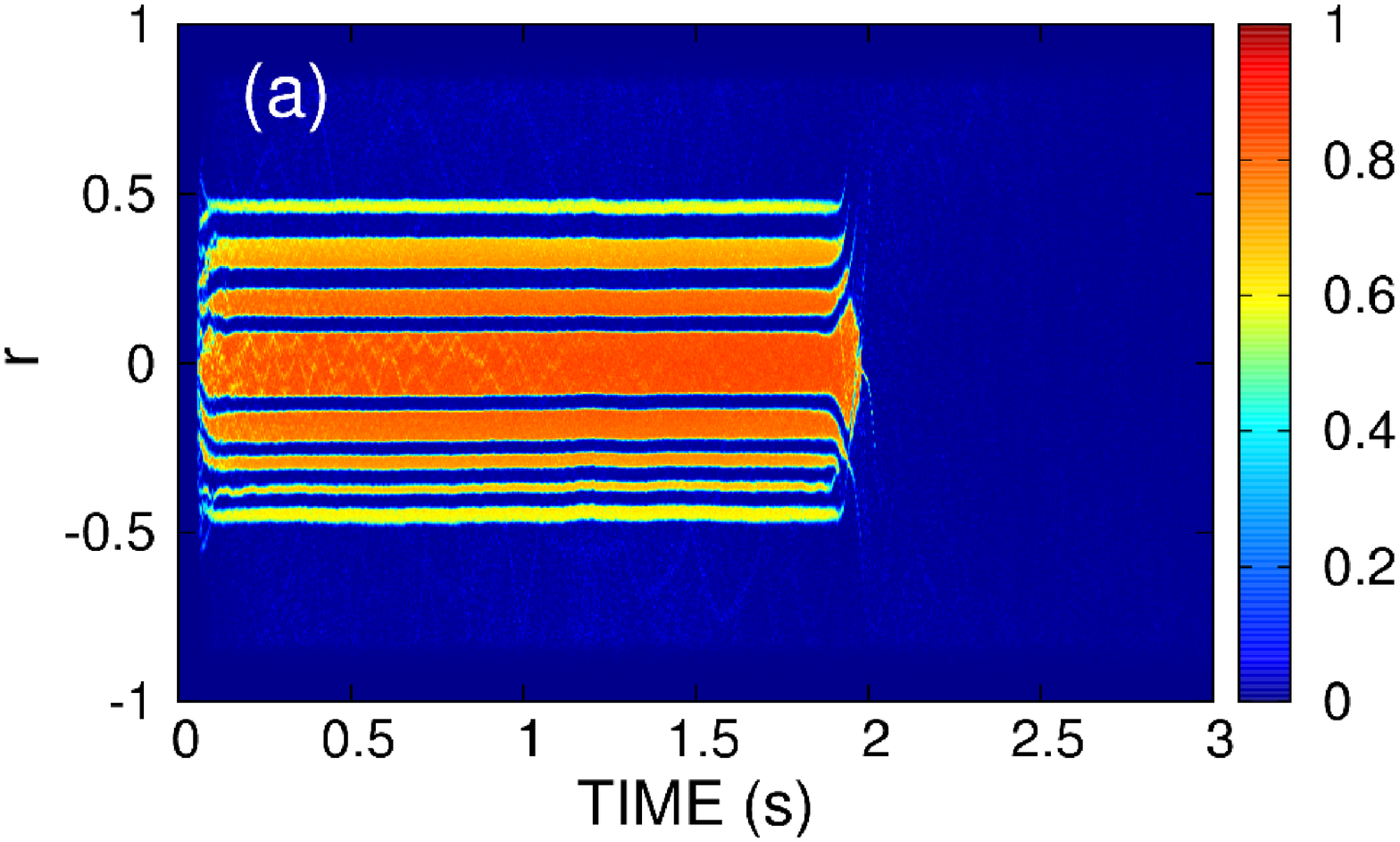}}
\put(-0,160){\includegraphics[width=0.16\textwidth,height=0.19\textheight,angle=-90]{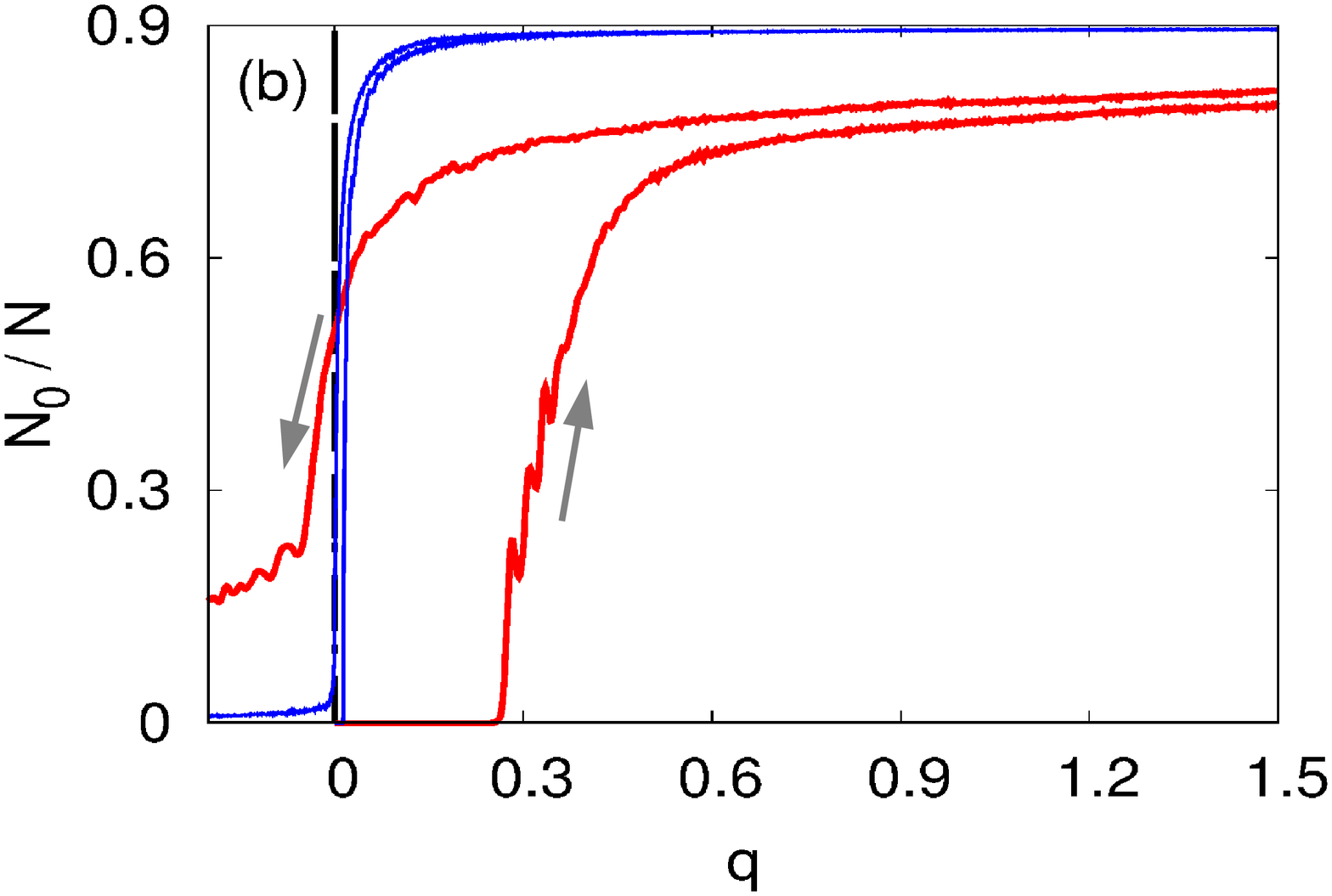}}
\end{picture}
\caption{(Color online) (a) Evolution of density of the $m_F=0$ component for $\tau_Q=1$s. 
(b), (c) and (d) show evolution of $N_0/N$ as a function of $q$ averaged over $50$ realizations for a harmonic trapping potential and $m=0.1$, $m=0.5$ and $m=0.95$ respectively. The thick red solid line corresponds to $\tau_Q=0.2$s, while the thin blue one to $\tau_Q=8$s. $q_1(0)$ (left) and $q_c(0)$ (right) are denoted by dashed and dot-dashed black lines respectively. The gray arrows indicate direction of evolution. The parameters are as follows: $N=2\times 10^6$, $\omega_\perp=2\pi \times 1000\,$Hz, $\omega= 2\pi \times 40\,$Hz, $\alpha=3$ and $t_{\rm max}=3\tau_Q$. 
}
\label{fig:fig4}
\end{figure}

In Figure~\ref{fig:fig3} we show examples of numerical simulation results for different fractional magnetizations in the large total atom limit $N=2\times 10^7$. Nucleation and growth of the $\rho_0$ phase and spin domains are clearly visible when the value of $q$ exceeds the critical value, see density profiles in Fig.~\ref{fig:fig3}(a). Several domains are created, not just two as predicted for the ground state, due to non-ideal adiabaticity of the quench. The number of domains decreases to two when increasing the quench rate $\tau_Q$. Initial oscillations in $N_0/N$ visible for shorter times in Fig.~\ref{fig:fig3}(b)-(d) result from spin-mixing dynamics~\cite{SMD1,SMD2,SMD3}. 
Indeed, for the smallest magnetization the hysteresis loop disappears as the quench time increases. On the other hand, the hysteresis loop of finite width is clearly visible and stable when the fractional magnetization tends to $1$, demonstrating the rate-independent hysteresis.

We also study the system enclosed in an external harmonic trapping potential ($\omega\ne 0$). One can show, based on the Thomas-Fermi and local density approximations, that the values of both $q_1$ and $q_c$ are space-dependent~\cite{ourPRAtrap}. 
By introducing the Thomas-Fermi unit $r=x/x_{\rm TF}$, where $x_{\rm TF}=[3c_0N/(2 \mu \omega^2)]^{1/3}$ is the Thomas-Fermi radius, the parameters $q$ of interest are
\begin{equation}
q_1(r) = \frac{1}{2}\left\{
\begin{array}{ccc}
\frac{(1-r_1^2)^2}{1-r^2}, & \,\,\,\,& |r|\in [0, r_1], \\
1-r^2, & \,\,\,\,& |r|\in [r_1,1] ,
\end{array}
\right.
\end{equation}
and
\begin{equation}\label{eq:Bc}
q_c(r) = (1-r^2) \left(  1- \sqrt{1 - \tilde{m}(r)^2} \right),
\end{equation}
where $r_1=(1-m)^{1/3}$ and the density of local magnetization is $\tilde{m}(r)=(1-r_1^2)/(1-r^2)$ for $r\in [0,r_1]$ and $\tilde{m}(r)=1$ otherwise.
The inhomogeneity, arising as a result of the external trapping potential brings in new physics. The width of the bistability region is not fixed but is space-dependent. 
Moreover, particular parts of the system undergo a phase transition at different times, which makes the bistability region additionally smeared out. 
The growth of the $\rho_0$ phase does not depend only on the reminiscence of the state history but is influenced much by the neighboring local phase due to tunneling of the local magnetization~\cite{ourPRAtrap}. As a consequence, the width of the hysteresis loop is not strictly connected to the width of the bistability area. In Fig.~\ref{fig:fig4} we show an example of the numerical simulation result for the evolution of density of the $m_F=0$ component and $N_0/N$ demonstrating the emergence of hysteresis.

\begin{figure}[]
\begin{picture}(0,160)
\put(-130,160){\includegraphics[width=0.17\textwidth,height=0.20\textheight,angle=-90]{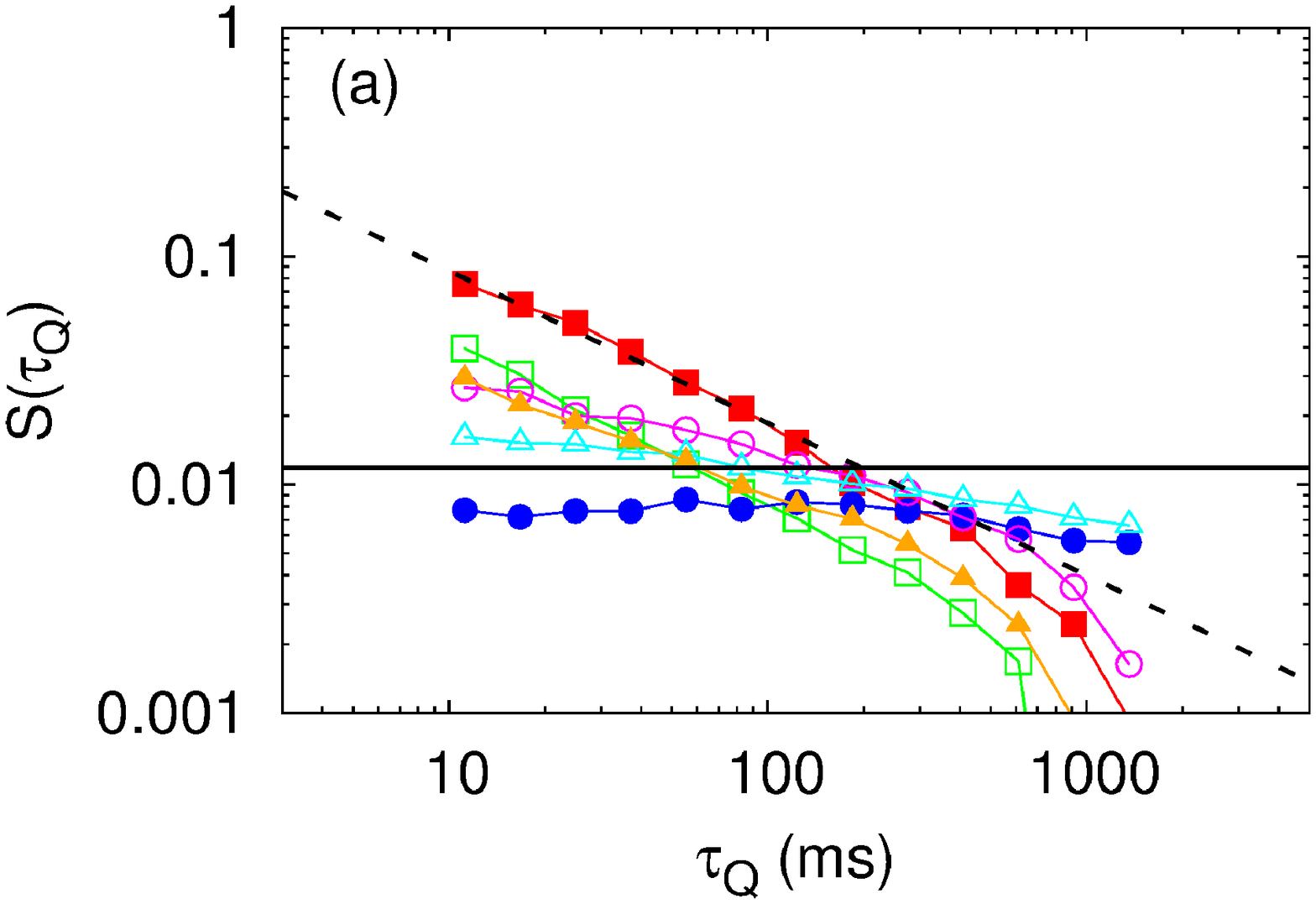}}
\put(-0,160){\includegraphics[width=0.16\textwidth,height=0.19\textheight,angle=-90]{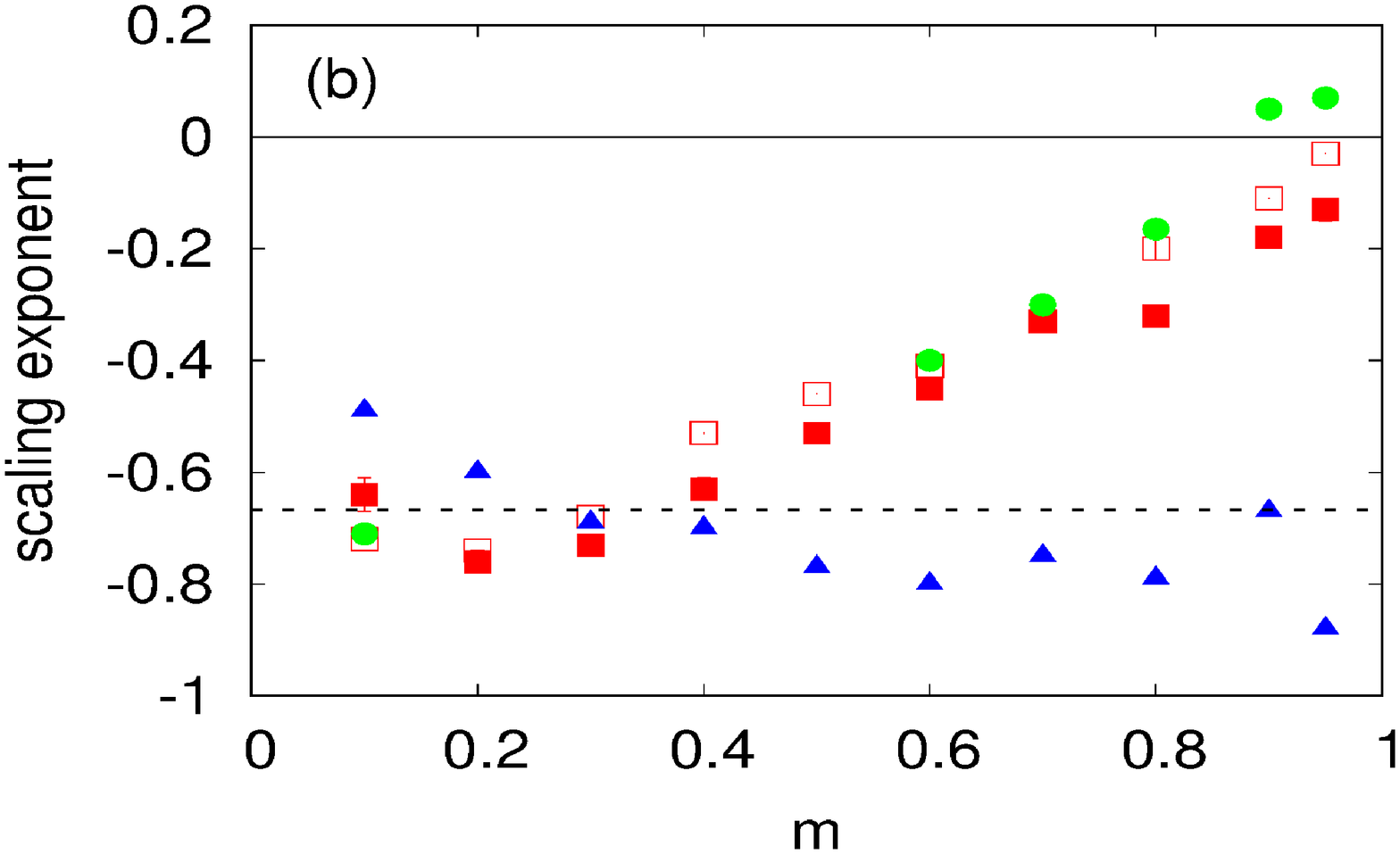}}
\put(-130,78){\includegraphics[width=0.17\textwidth,height=0.20\textheight,angle=-90]{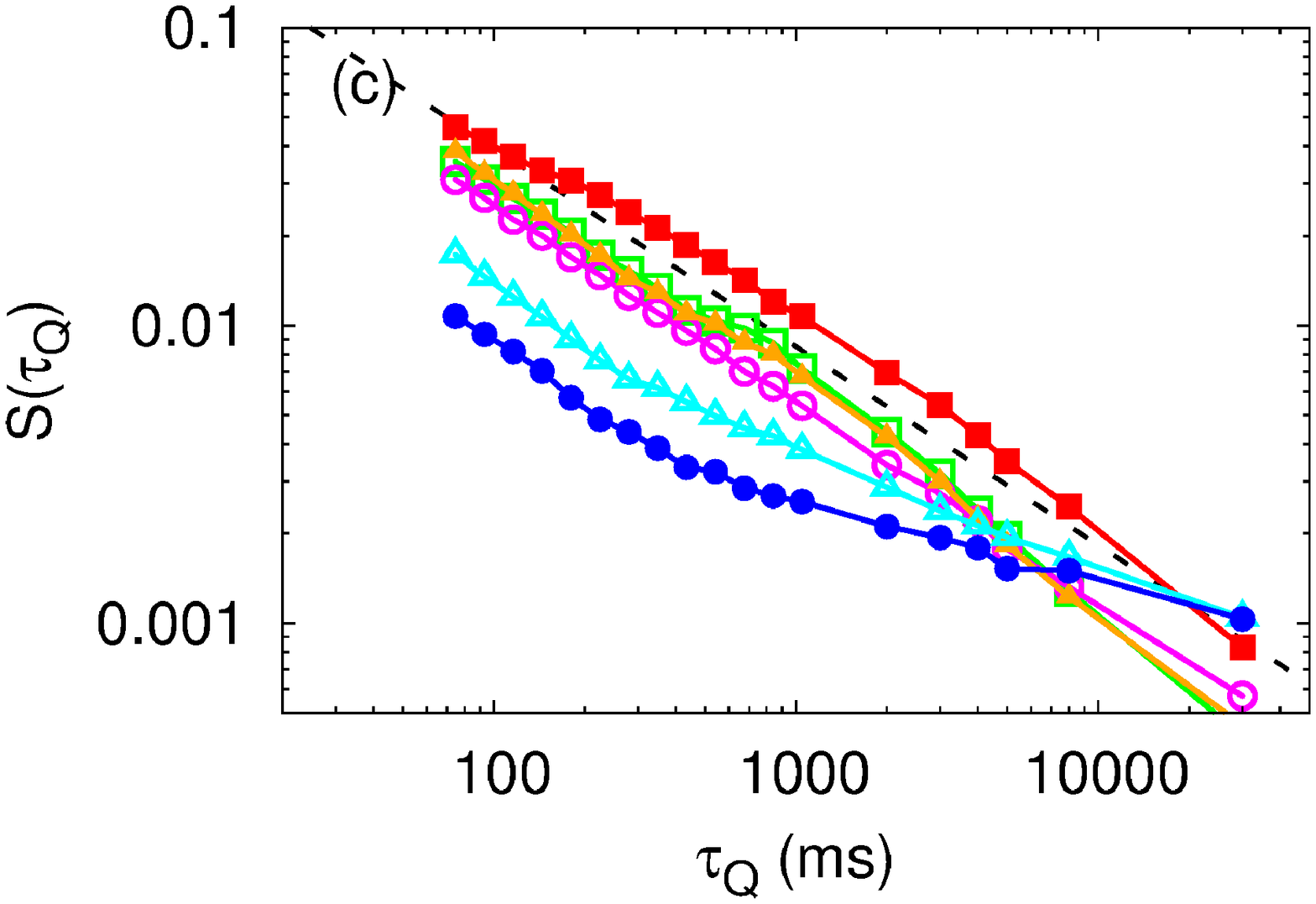}}
\put(-0,78){\includegraphics[width=0.16\textwidth,height=0.19\textheight,angle=-90]{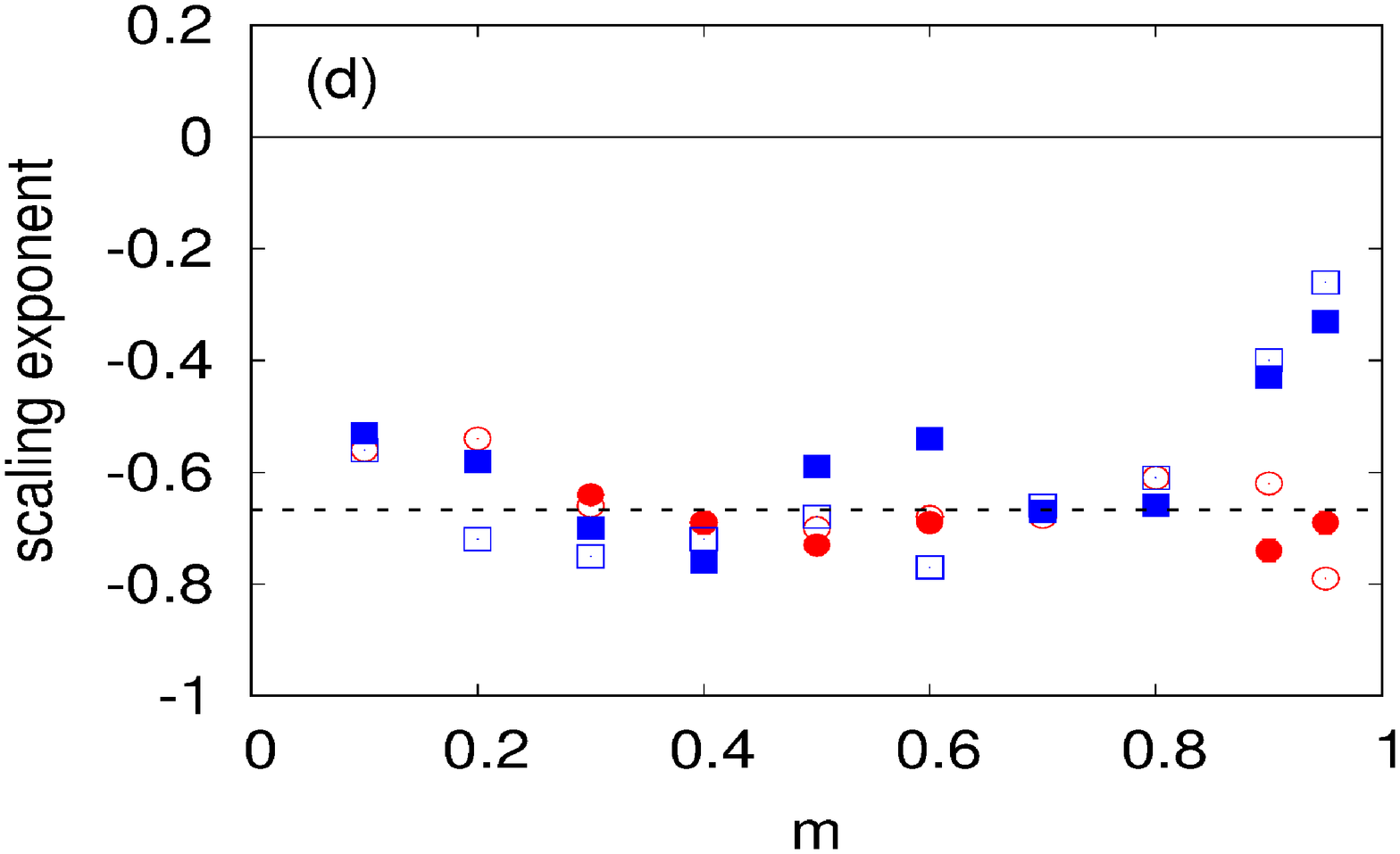}}
\end{picture}
\caption{
(Color online) (a) and (c) scaling of the hysteresis loop area $S(\tau_Q)$ with $\tau_Q$ for $m=0.1$ (filled squares), $m=0.3$ (open squares), $m=0.5$ (filled triangles), $m=0.7$ (open circles), $m=0.9$ (open triangles) and $m=0.95$ (filled circles) for the box and harmonic trap potentials, respectively. The dashed line shows $S(\tau_Q) \propto \tau_Q^{-2/3}$, while the solid line shows the value $(1-m)(q_c-q_1)$ for $m=0.95$. 
(b) and (d) show scaling exponent of the hysteresis area versus fractional magnetization $m$ obtained numerically by fitting a linear function to logarithms of numerical data presented in (a) and (c), respectively.
The total number of atoms in (b) is $N=2\times 10^7$ (squares), $N=6\times 10^5$ (circles) and $N=10^4$ (triangles).
The total number of atoms in (d) is $N=2\times 10^6$ and the circles (squares) mark the scaling exponent of 
the hysteresis area for $\tau_Q<1$s ($\tau_Q>1$s). 
In (b) and (d) open symbols correspond to $\alpha=1.5$, while filled ones to $\alpha=3$.  Notice, error bars are shown for $N=2\times 10^7$ in (b) and for $N=2\times 10^6$ in (d), but they are of the order of the symbol size.
}
\label{fig:fig5}
\end{figure}

In Fig.~\ref{fig:fig5}(a) we show scaling of the hysteresis area $S$ versus ramp times $\tau_Q$ for the box-like potential ($\omega=0$). The two limiting cases, $m\to 0$ and $m\to 1$, are clearly visible. Interestingly, even for intermediate fractional magnetizations the hysteresis area $S$ is subject to the scaling law but with a different exponent. We gather the resulting scaling exponents versus fractional magnetization in Fig.~\ref{fig:fig3}(b). While the results for the largest and moderate atom numbers follow our predictions, the results for the smallest atom number ($N=10^4$  marked by triangles) are different. This is because the widths of domain walls increase when $N$ decreases, and the energy of the domain wall cannot be neglected in the derivation of $q_1$. In other words, the effect of finite size of the system increases the value of $q_1$ up to $q_c$, see Fig.~\ref{fig:fig1}. 
Consequently, the width of the bistability region tends to $0$ and the hysteresis phenomenon becomes rate-dependent even for large fractional magnetizations, as demonstrated in Fig.~\ref{fig:fig5}(b). The resulting scaling exponent in the low density regime follows the KZ theory~\cite{ourPRB} slighty modified by the phase ordering kinetics process~\cite{Bray1994}.

The case of a trapped system ($\omega\ne 0$) is qualitatively different because the scaling of the hysteresis loop area exhibits double law for macroscopic magnetizations $m\to 1$, as illustrated in Fig.~\ref{fig:fig5}(c) for $N=2\times 10^6$. The KZ theory for the trapped case in the local density approximation~\cite{ourPRAtrap} gives $\hat{S}\propto \tau_Q^{-2/3}$ which is confirmed by numerical calculations presented in Fig.~\ref{fig:fig5}(d) for small $\tau_Q$ only. The numerical results deviate from the KZ theory predictions for macroscopic magnetizations in the adiabatic quench times limit.

While performing the experiment with the largest number of atoms ($N=2\times10^7$) appears to be unrealistic, in part beacause of strong two- and three-body losses, we find a regime of parameters in which the hysteresis and scaling laws may be observed, as shown by green dots in Fig.~\ref{fig:fig5}(b). We propose to use a moderate number of atoms $N=6\times10^5$, and a tight transverse confinement of $\omega_\perp=2\pi \times 2 800\,$Hz, which allows one to avoid transverse excitations. The latter tight confinement requirement may be reduced down to $\omega_\perp=2\pi \times 100\,$Hz in the non-polynomial Gross-Pitaevskii regime~\cite{PRA 65 043614 (2002)}, as long as the ratio $c_2 n / \hbar \omega_\perp$ is small, which assures the absence of transverse {\it spin} excitations. The spatial extent of the condesate in the longitudinal direction must be large enough so that sufficiently many domains are observed ($L>600\mu$m), which appears to be within the reach of state-of-the-art experiments~\cite{WolfvonKlitzing_HPM2017_Conference}. With these parameters, and typical linear densities of $10^{14}$ atoms/cm$^3$, the lifetime of the condensate due to one-, two- and three-body losses may be estimated at around 20s~\cite{Ketterle_NaF2}, which should allow one for the observation of rate-independent hysteresis predicted here.

In summary, an antiferromagnetic spinor condensate exhibits hysteresis controlled by the magnetic field, which may be practical in its further applications.
We investigated hysteresis during the phase transition, showing that it is rate-independent for a homogeneous system in the limit of large magnetizations only. In all other cases, the hysteresis is rate-dependent and the area of its loop is subject to the universal scaling law which we explained based on the KZ theory. 

We acknowledge support from the National Science Center Grants DEC-2015/18/E/ST2/00760, DEC-2015/17/B/ST3/02273, and DEC-2015/17/D/ST2/03527.



\begin{thebibliography}{99}
\bibitem{foot1}
The width of the domain wall is of the order of the spin healing length, and the condition $\xi_s\ll L$ allows one to neglect the energy of the domain walls during derivation of $q_1$ and $q_2$~\cite{{Matuszewski_PS},{ourPRB}}.
\end{thebibliography}
\end{document}